\documentstyle[12pt,epsf]{article}

\newcommand{\beq}{\begin{equation}}
\newcommand{\eeq}{\end{equation}}

\newcommand{\bea}{\begin{eqnarray}}
\newcommand{\eea}{\end{eqnarray}}

\title{
Precision study of positronium and precision tests\\ 
of the bound state QED
}

\author{Savely G. Karshenboim\thanks{E-mail: sek@mpq.mpg.de}\medskip\\
D. I. Mendeleev Institute for Metrology, 198005 St. Petersburg, 
Russia\\
Max-Planck-Institut f\"ur Quantenoptik, 75748 Garching, Germany
}

\date{}

\begin{document}

\maketitle
\begin{abstract}
Despite its very short lifetime positronium provides us with a number 
of accurate tests of the bound state QED. In this note 
a brief overview of QED theory and precision experiments on the  
spectrum and annihilation decay of the positronium atom is presented. 
Special attention is paid to the accuracy of theoretical predictions.
\end{abstract}

\section{Introduction}

Positronium is a very specific two-body atomic system, which  
in some respects is similar to hydrogen, but some its features  
are quite different from those of the hydrogen atom. In particular, 
positronium is an unstable atom and the lifetime of its ground state 
is about $10^{-7}$ or $10^{-10}$~sec depending on its spin. 
Because of that any experiments with 
positronium are much more complicated than those with hydrogen and 
considerably less precise. In this brief overview of positronium 
studies we will try to demonstrate that despite the existing experimental 
problems positronium is worth studying. 

The theory which describes simple atomic systems is bound state Quantum 
Electrodynamics (QED) and the questions we are trying to answer in this 
paper are
\begin{itemize} 
\item Why do we have to study and test {\em bound state QED}?
\item What are {\em essential problems} of present-day bound state QED?
\item What is {\em specific} with {\em positronium} tests of QED?
\end{itemize} 

Bound state QED is Quantum Electrodynamics for the bound states and a  
problem of the bound states is a complicated one even in the case of 
classical physics. The bound state QED theory has been mainly developed 
to describe two-body and three-body atoms. Even in the case of free 
mass-shell particles, QED as a theory of interactions between leptons 
(electrons and muons) and photons is indeed incomplete. It faces a lack of 
pure QED description of the nucleon structure, hadronic vacuum polarization 
and other hadronic effects. Such effects are unavoidable while calculating 
the spectrum of hydrogen, muonium and most of other simple atoms \cite{icap}. 
For some applications the weak interaction is also involved into 
calculations, but in such a case it can often be done {\em ab initio}, 
while strong interaction effects require model-dependent evaluations, 
experimental data and phenomenological approaches.

QED cannot predict a number to be compared with some experimental value, 
producing instead some expressions which need appropriate values of 
fundamental constants. These constants as well as a number of auxiliary 
nuclear parameters should be determined experimentally.
Measurements of the constants and parameters are apparently a problem 
beyond QED. For most of the so called tests of bound state QED the 
accuracy is usually limited by factors related to non-QED phenomena and 
the actual goal of the study is rather to determine the constants and 
nuclear properties. 

A progress of theory is required if one intends to separate nuclear and 
QED effects in order to determine the nuclear parameters (like e.g. the 
proton charge radius). It is also 
needed to improve the accuracy of the fundamental constants obtained this way 
(like the Rydberg constant, the fine structure constant $\alpha$ etc.). 

Presently QED calculations have some essential problems to be solved. Those 
are related to the bound-state effects of atomic states. Theory involves a 
number of small parameters such as the fine structure constant $\alpha$, 
the Coulomb strength $Z\alpha$, the electron-to-nucleus mass ratio $m/M$ etc. 
Because of that no calculation can be exact and some effective 
expansion over small parameters is involved.
Higher order corrections become more and more complicated and at any time 
there are some which cannot be calculated because of their complicated 
analytic structure, huge amount of diagrams, numerous lengthy computations 
etc. That faces the basic problem of real QED, {\em how to estimate the 
corrections we cannot calculate?} 
The problem exists for both QED of a free particle and the bound state QED. 
However, in the case of the bound state problem the expansion over $Z\alpha$ 
and $m/M$ is not analytic and involves large logarithmic factors and big 
numerical coefficients. The front line of the today calculations
is the study of three kinds of corrections:
\begin{itemize} 
\item the higher-order two-loop corrections of order 
$\alpha^2(Z\alpha)^6m$ crucial for theory of the hydrogen Lamb shift; 
\item radiative-recoil corrections of order $\alpha(Z\alpha)^6m^3/M^2$
which are critical contributions to the muonium hyperfine structure (HFS); 
\item pure recoil corrections of order $(Z\alpha)^7m^3/M^2$ which are 
to be calculated to improve the accuracy of the muonium HFS. 
\end{itemize} 

Since for most of simple atoms any precision test of the bound state of QED 
involves numerous effects beyond QED, it should be of great interest to 
develop some test which will be related to a pure QED quantity. Positronium 
is one of very few atoms which offer such an opportunity. Few features of 
positronium make it a very useful system to test the bound state QED. 
Like muonium, it is a pure leptonic atom with no nuclear structure. However, 
in contrast to muonium, the electron-to-nucleus mass ratio is unity and 
that allows to study accurately higher order recoil effects performing 
relatively low accuracy experiments. E.g. a study of the 
recoil contributions to the positronium HFS in order 
$\alpha(Z\alpha)^6m^3/M^2$ and $(Z\alpha)^7m^3/M^2$ 
with the same precision as for muonium ($m/M\sim1/200$) 
requires a fractional accuracy two orders of magnitudes 
below that for muonium. The relatively low accuracy offers a possibility to 
perform all calculations and measurements at the level above 1 ppm and 
to separate QED problems from a problem of the 
determination of the fine structure constant. The latter is related to 
the level of accuracy significantly below 1 ppm. Absence of any 
serious problems with the accurate determination of fundamental constants is 
a significant advantage of positronium studies.

One further advantage is a great variety of quantities which can in 
principle be investigated with high accuracy. A number of spectral values 
can be measured ($1S-2S$ interval, fine structure at $n=2$ and $1S$ HFS and 
some others) and they are {\em not 
sensitive} to any new physics beyond the Standard Model. That is a direct 
result of the lightness of the nuclear mass (positronium mass) which does 
not allow any high momentum in any virtual effects. However, physics 
beyond the Standard model can be studied within non-spectral experiments 
with positronium, particularly in the case of exotic decay modes. A 
possibility to clearly separate quantities insensitive and 
possibly sensitive to the new physics is another advantage of 
the positronium studies.

\section{Present status of precision physics of positronium}

About ten years ago the experimental accuracy was essentially better than 
the theoretical one for most of the quantities under study. The last 
decade delivered a great theoretical improvement. We have collected most 
of important theoretical references in Table~\ref{PsRef} and will briefly
describe the recent progress in this section. In the case of the positronium 
spectrum the critical contributions to the $1S$ HFS and to the $1S-2S$ 
interval are of the order $\alpha^6m$ and their calculation was completed 
some years ago. Some results were obtained numerically and later improved 
by analytic calculations (see e.g. Refs.~\cite{pk1,cmy2}). However, minor 
shifts (even equal to few standard deviations of the numerical integration) 
did not affect the final results because considerably higher  
uncertainty arose because of another effect related to 
uncalculated higher-order terms. It was an exception related to the 
so called pure recoil spin-dependent $\alpha^6m$ contribution, calculated by 
several authors with contradicting results. The later analytic calculations 
\cite{cmy2} confirmed the numerical result from Ref.~\cite{pach}. 

\begin{table}[tbh]
\begin{center}
\begin{tabular}{cccc}
\hline
Value & Ref. to & Ref. to & Ref. to \\
& $\alpha^6m$ & $\alpha^7m \ln^2\alpha $& $\alpha^7m\ln\alpha$\\
& or $\alpha^2\Gamma^0$ &$\alpha^2\ln^2\alpha\Gamma^0$ & $\alpha^2\ln\alpha\Gamma^0$\\ 
\hline 
$1S-2S$ & \protect\cite{pk1} &\protect\cite{pk2}  & unknown \\
fine structure &\protect\cite{pk1} &\protect\cite{pk2}  & unknown \\
$1S$ HFS &\protect\cite{ahfs} &\protect\cite{93}  &\protect\cite{hfslog} 
\\
\hline
$\Gamma({\rm p\!-\!Ps})$ &\protect\cite{cmy1} &\protect\cite{93}  &\protect\cite{declog} 
\\
$\Gamma({\rm o\!-\!Ps})$ &\protect\cite{afs} &\protect\cite{93}  &\protect\cite{declog} 
\\
\hline
\end{tabular}
\end{center}
\caption{\label{PsRef} References to recent progress in positronium theory. 
The contributions to the spectrum are classified by the electron mass, 
while those to the decay are presented in units of the leading contribution 
$\Gamma^0$. 
}
\end{table}

The uncertainty of any theoretical calculation is determined by a 
possible value of unknown higher-order corrections which are expected to 
have large coefficients. There is a number of corrections enhanced by a 
big double logarithmic factor $\ln^2\alpha\simeq 24$ \cite{93} and 
the higher-order terms should  be studied to better understand the accuracy 
of theory. It is also necessary to investigate terms beyond the leading 
logarithms. The higher-order terms have been known only in part.

In the case of decay theoretical problems are related to corrections of 
the relative order $\alpha^2$ which were calculated only recently. The 
calculation of the $\alpha^3$ contributions is now in progress and they 
are known in the logarithmic approximation.

\begin{table}[tbh]
\begin{center}
\begin{tabular}{lc}
\hline
Quantity & Prediction \\
\hline 
$\Delta\nu(1S-2S)$  & 1\,233\,607\,222.2(6) MHz\\
$\Delta\nu_{HFS}(1S)$ & 203\,391.7(5) MHz\\
\hline
$\Delta\nu(2^3S_1-2^3P_0)$  & 18\,498.25(9) MHz \\
$\Delta\nu(2^3S_1-2^3P_1)$  & 13\,012.41(9) MHz\\
$\Delta\nu(2^3S_1-2^3P_2)$  & 8\, 625.70(9) MHz\\
$\Delta\nu(2^3S_1-2^1P_1)$  & 11\,185.37(9) MHz\\
\hline 
$\Gamma ({\rm p\!-\!Ps})$& 7\,989.32(2) $\mu$s$^{-1}$ \\
$\Gamma ({\rm o\!-\!Ps})$& 7.040\,07(2) $\mu$s$^{-1}$ \\
\hline
\end{tabular}
\end{center}
\caption{\label{TPsTh} Theoretical predictions for positronium.}
\end{table}

We collected all theoretical predictions in Table~\ref{TPsTh}. The 
results were published and presented in different compilations. What we 
would like to underline here is our {\em estimation of uncertainty}. 
For most of the quantities not only the leading logarithmic corrections 
(e.g. in the case of spectrum that is 
$\alpha^7m\ln^2\alpha$) are known, but also the next-to-leading term 
($\alpha^7m\ln^2\alpha$). However, that cannot reduce the uncertainty because
the leading term originates from a single source and its magnitude is 
characteristic of the correction, while the next-to-leading term used is 
a result of cancelation between different contributions and can be sometimes 
small. But that smallness is misleading and the constant following the 
single logarithm is not small. Our estimation of the uncertainty is based 
on a value of the double logarithmic term in any case (see Ref.~\cite{cek} 
for more detail). 

\section{Summary of positronium study}

\begin{figure}[tbh]
\epsfxsize=4.5cm
\centerline{\epsfbox{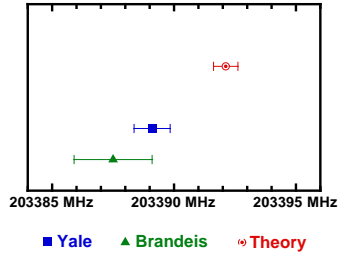}}
\caption{\label{Fhfs} $1S$ HFS in positronium}
\end{figure}

\begin{figure}[tbh]
\epsfxsize=4.5cm
\centerline{\epsfbox{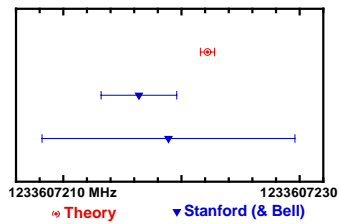}}
\caption{\label{F1s} $1S-2S$ interval in positronium}
\end{figure}

Studies of the spectrum and decay rates of positronium provide 
us with a number of the strong tests of bound state QED, some of which 
are among the most accurate. Some theoretical predictions from 
Table~\ref{TPsTh} can be compared with accurate experimental data, 
a review of which can be found in Refs.~\cite{conti,ley}. The most 
accurately measured spectroscopic data are related to the $1S-2S$ interval 
(see Fig.~\ref{F1s}) and to the ground state HFS (see Fig.~\ref{Fhfs}). 
There are some minor discrepancies between experimental and theoretical data.

\begin{figure}[tbh]
\epsfxsize=6cm
\centerline{\epsfbox{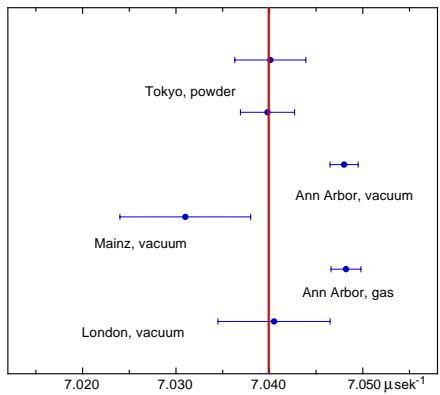}}
\caption{\label{Fdop} Decay of orthopositronium}
\end{figure}

\begin{figure}[tbh]
\epsfxsize=4.5cm
\centerline{\epsfbox{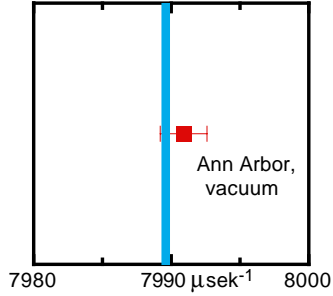}}
\caption{\label{Fdpp} Decay of parapositronium}
\end{figure}

The current experimental situation with the orthopositronium decay 
(see Fig.~\ref{Fdop}) is not acceptable at all. The main problem is 
inconsistency of various experiments. Note that we included a new result 
from Tokyo \cite{eopsj2} and corrected a gas value from Ann Arbor 
according to the preliminary analysis in Ref.~\cite{conti}. In contrast, 
theory and experiment are in a fair agreement for the parapositronium decay 
(see Fig.~\ref{Fdpp}). References to experimental results can be found 
in Refs.~\cite{conti,ley}. These two papers also review experiments on the 
fine structure in positronium performed at $2^3S_1-2P$ intervals which 
were less accurate than experiments at $1S$ HFS and $1s-2S$ intervals.

\section{What we can learn from positronium?}

To understand the role of positronium tests of the bound state QED,
let us have a look at crucial contributions which can be studied in
different tests. They are collected in Table~\ref{TallQED}. One sees that a 
spectroscopic study with positronium allows a few high-precision tests 
for higher-order recoil effects needed for other atoms. Positronium theory 
is a good training field for a number of other atomic systems. First of all, 
let us note that the positronium is in some sense an only `true' two-body 
atom among all hydrogen-like QED systems. 
In the case of hydrogen, muonium and others most of the calculations 
are performed for an electron bound by an external Coulomb field, and only 
for a few corrections the two-body effects are significant. In contrast, 
for positronium, the two-body phenomena are essential from the very 
beginning of any calculation. Due to that positronium studies have been 
important for understanding higher-order QED effects in helium, a system 
which can mainly be treated as a system of two electrons in an external field.

\begin{table}[tbh]
\begin{center}
\begin{tabular}{lc}
\hline
Value & Order \\
\hline 
hydrogen (gross structure) & $\alpha^8m$ \\
hydrogen (fine structure)  & $\alpha^8m$ \\
hydrogen (Lamb shift)      & $\alpha^8m$ \\
He$^+$ (Lamb shift)        & $\alpha^8m$ \\
nitrogen (fine structure)  & $\alpha^8m$ \\
$^3$He$^+$ HFS             & $\alpha^8m^2/M$,\\
& $\alpha^7m^3/M^2$\\
muonium HFS                & $\alpha^8m^2/M$,\\  
& $\alpha^7m^3/M^2$\\
positronium HFS            & $\alpha^7m$ \\
positronium (gross structure)      & $\alpha^7m$ \\
positronium (fine structure)       & $\alpha^7m$ \\
parapositronium (decay rate)       & $\alpha^7m$ \\
orthopositronium (decay rate)      & $\alpha^8m$ \\
parapositronium ($4\gamma$ branching)  & $\alpha^8m$ \\
orthopositronium ($5\gamma$ branching) & $\alpha^8m$ \\
\hline
\end{tabular}
\end{center}
\caption{\label{TallQED} Crucial QED contributions for most important tests 
of the bound state QED. Tests related to positronium are exact in $m/M$ 
since $m/M=1$.}
\end{table}

Development of the bound state QED is also fruitful for a better 
understanding of hadronic systems, like deuteron (proton-neutron system) and 
mesons (quark-antiquark systems). In both cases a consideration of 
the  $m/M$  value close to unity is of particular interest.

Successful development of theory during the last decade has made it more 
accurate than the experiment and we hope that some progress from the 
experimental side will come. That is in particular related to the fine 
structure at $n=2$.

\end{document}